\documentclass[aps,twocolumn,english,aip,superscriptaddress,amssymb,showpacs,preprintnumbers,amsmath,amssymb,superscriptaddress]{revtex4-1}

\usepackage{graphicx}
\usepackage{dcolumn}
\usepackage{bm}
\newcommand{\nc}{\newcommand}
\nc{\be}{\begin{equation}}
\nc{\ee}{\end{equation}}
\nc{\bea}{\begin{eqnarray}}
\nc{\eea}{\end{eqnarray}}
\nc{\bean}{\begin{eqnarray*}}
\nc{\eean}{\end{eqnarray*}}
\nc{\mb}{\mbox}
\nc{\rnc}{\renewcommand}
\nc{\vk}{\mb{\bf k}}
\nc{\vp}{\mb{\bf p}}
\nc{\vn}{\mb{\bf n}}
\nc{\vq}{\mb{\bf q}}
\nc{\rr}{\mb{\bf r}}
\nc{\vz}{\hat {\mb{\bf z}}}
\nc{\vj}{\mb{\boldmath$j$}}
\nc{\vg}{\mb{\boldmath$g$}}
\nc{\x}{\mb{\boldmath$x$}}
\nc{\A}{\mb{\boldmath$A$}}
\nc{\va}{\mb{\boldmath$a$}}
\nc{\vs}{\mb{\boldmath$\sigma$}}
\nc{\vpi}{\mb{\boldmath$\pi$}}
\nc{\nab}{\nabla}
\nc{\X}{\sf x}

\begin{document}
\title{Spin superfluid Josephson oscillator }

\author{Yizhou Liu}
\thanks{yliu062@ucr.edu}
\affiliation{Department of Electrical and Computer Engineering, University of California, Riverside, California 92521, USA}

\author{Igor Barsukov}
\affiliation{Department of Physics and Astronomy, University of California,
Riverside, California 92521, USA}

\author{Ilya Krivorotov}
\affiliation{Department of Physics and Astronomy, University of California,
Irvine, California 92697, USA}

\author{Yafis Barlas}
\thanks{yafisb@gmail.com}
\affiliation{Department of Electrical and Computer Engineering, University of California,
Riverside, California 92521, USA}

\author{Roger K. Lake}
\thanks{rlake@ece.ucr.edu}
\affiliation{Department of Electrical and Computer Engineering, University of California,
Riverside, California 92521, USA}

\begin{abstract}
The magnetic analogue of the Josephson effect can be exploited to develop a new class of nano-spin
oscillators that we denote as spin superfluid Josephson oscillators.
Such a device, consisting of two exchange coupled easy-plane metallic 
ferromagnets separated by a thin normal metal spacer, is proposed and analyzed.
A spin chemical potential difference drives a $2\pi$ precession of the in-plane magnetization of each ferromagnet.
The $2 \pi$ precession angle gives maximum values of the giant magnetoresistance, 
resulting in large output power compared to conventional spin Hall oscillators.
An applied ac current results in a time-averaged magnetoresistance with Shapiro-like steps. 
The multistate mode-locking behavior exhibited by the spin Shapiro steps may be explored for applications in neuromorphic computing.
As an experimental characterization method, 
electrical measurements of spin superfluid Josephson junctions can provide additional signatures of spin superfluidity.

\end{abstract}

\pacs{}

\maketitle

%
%
%

\section{Introduction}
One of the goals of spintronics is the transport of spin with minimal losses.
%
%
In magnetic insulators with spontaneously broken U(1) symmetry, 
almost dissipationless longitudinal spin transport can result from the 
collective excitations of the ground state~\cite{HalperinHohenberg_69,sonin_review}.
Such a spin current decays algebraically
over distance~\cite{sonin_review,takei_superfluid_2014}, 
which, in comparison to the exponential decay of 
magnon mediated spin currents~\cite{SZhangmagnondrag}, makes it attractive for long distance spin transport.
The spin current in this state is carried by a metastable static spin spiral texture 
with $2 \pi$ phase winding, much like phase slips in superfluids.
This spin superfluid transport in easy-plane ferromagnets (FMs) 
and antiferromagnets (AFMs)~\cite{MacDSSF1,sun_spin_polarized_2013, takei_superfluid_2014, MacDSSF2, takei_superfluid_afm} 
is formally similar to mass or charge currents in superfluids or superconductors. 
Experimental evidence consistent with spin superfluid transport in 
a graphene quantum Hall antiferromagnet was recently reported~\cite{stepanov_long_distance_2018}.
A natural extension of this idea is the magnetic analogue of the Josephson 
effect~\cite{josephson_possible_1962,2004_Nogueira_Spin_Josephson_EPL,Schilling_Ann_Phys,2018_Long_Josephson_YTserkov_arxiv}.

This spin superfluid Josephson effect can be harnessed to develop a new class of nano-spin oscillators
that we denote as spin superfluid Josephson oscillators (SSJOs). 
Spin oscillators convert dc electric current into non-linear magnetization precession~\cite{katine_current-driven_2000,sun_current-driven_1999,tsoi_excitation_1998, kiselev_microwave_2003, rippard_direct-current_2004, krivorotov_time-domain_2005}, 
which can be detected from the magnetoresistance.
Persistent magnetization oscillations can also be induced in ferromagnetic insulators (FMI) and 
antiferromagnetic insulators (AFMI).
In these systems, a pure spin current, injected via the spin Hall effect (SHE) from
an adjacent heavy metal (HM) \cite{collet_generation_2016, hamadeh_full_2014}, 
causes magnetization dynamics in the FMI or AFMI layer.
Recently, a SSJO was proposed using
exchange coupled AFMIs \cite{liu_spin-josephson_2016}.
While AFMIs have low damping, negligible dipolar coupling, and THz frequencies,
the a.c. output power of the AFMI SSJO is low.
Here, we demonstrate an easy-plane metallic FM SSJO
that exhibits large output power compared to that of conventional spin oscillators.

\begin{figure}
\begin{center}
\includegraphics[width=3in,height=2.2in]{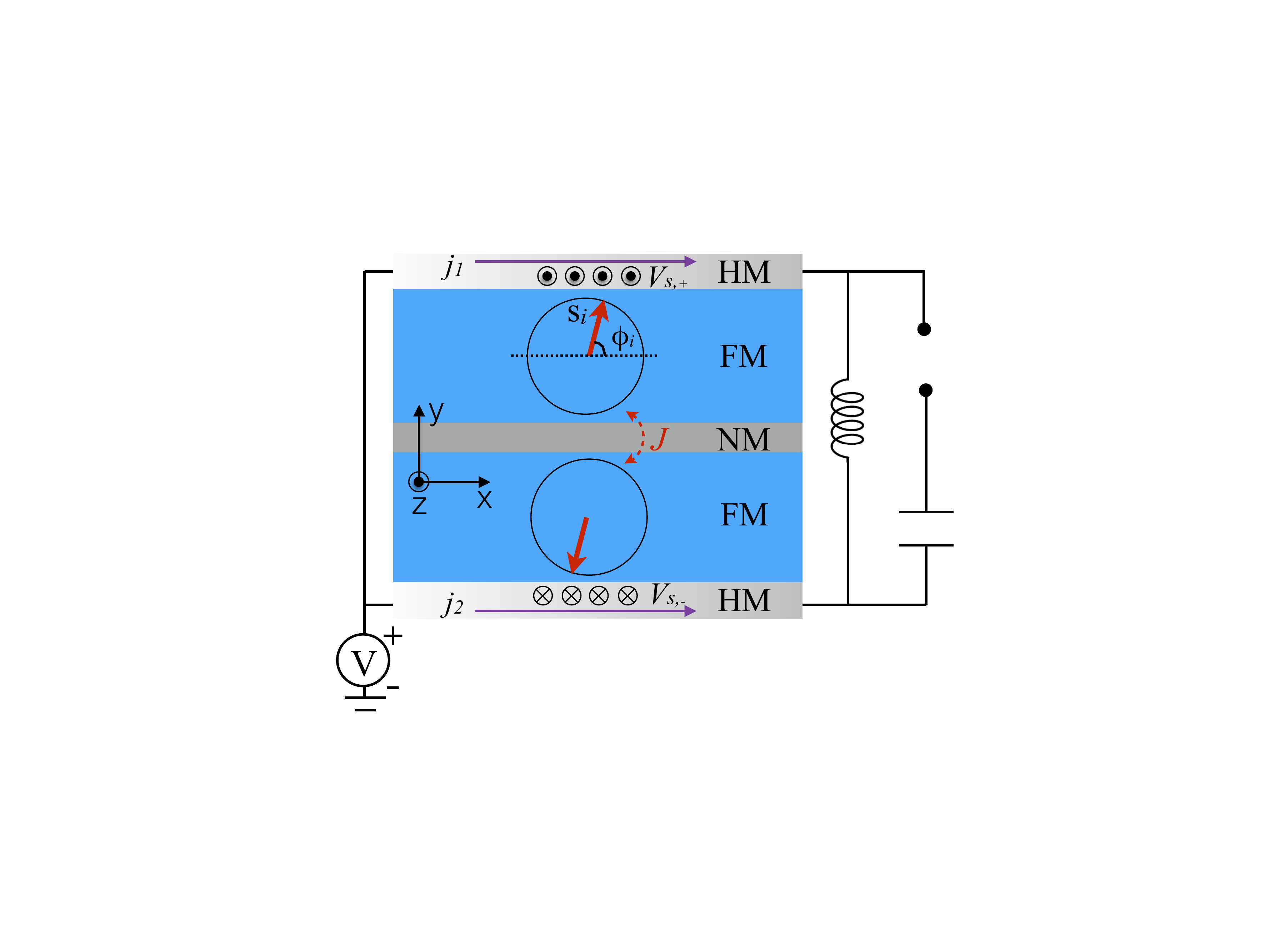}
\caption{Schematic diagram of the SSJO. 
Two easy-plane FMs, with magnetization in the $x$-$y$ plane, are separated by a thin nonmagnetic metallic (NM) spacer. 
The FMs are exchange coupled via an antiferromagnetic type interlayer exchange coupling $J$. 
The FM junction is sandwiched by two identical heavy metals.}
\label{fig:structure}
\end{center}
\end{figure}

The SSJO illustrated in Fig.~\ref{fig:structure} consists of two easy-plane FMs 
with an antiferromagnetic interlayer exchange coupling.
This is the magnetic analogue of a $\pi$-phase Josephson junction.
Electrical current along the x-direction in the HM generates a transverse spin current across the junction.
Using both the Landau-Lifshitz-Gilbert (LLG) equation and micromagnetic simulations, we show that this current drives steady-state spin oscillations within the junction with a frequency determined by the applied current and the parameters of the 
spin superfluid Josephson (SSJ) junction.
These spin oscillations result from $ 2 \pi$ rotations of the in-plane magnetization of the easy-plane FMs.

The SSJ produces a novel magnetoresistance (MR) effect, hereafter called SSJ-MR, which is the only contribution in insulating devices.
In metallic systems, the MR signal induced by the $2 \pi$ phase rotation has an additional 
contribution from the giant magnetoresistance (GMR) effect.
The global 2$\pi$ precession of the in-plane magnetization results in large output power of the SSJO.
When connected with an ac current, the time-averaged SSJ-MR exhibits Shapiro steps as a function of the applied current's dc component.
The SSJ-MR exhibits a step when the ac driving frequency matches the characteristic frequency of the SSJO.

\section{Theoretical Model}
The SSJO shown in Fig. \ref{fig:structure}
consists of two HM contacts and two easy-plane FMs separated by a thin non-magnetic spacer.
The Hamiltonian for this system can be expressed as,
\begin{eqnarray}
\label{eq_Hamiltonian}
\mathcal{H} &=& \frac{1}{\mathcal{V}}\int d\textbf{r} \sum_{i= \pm } \frac{1}{2} [ A \big(\nabla \textbf{s}_{i}(\textbf{r}) \big)^2 + K s_{i}^z(\textbf{r})^2  \\ \nonumber 
&+& J \textbf{s}_{i}(\textbf{r})\cdot \textbf{s}_{-i}(\textbf{r}) ],
\end{eqnarray}
where $\mathcal{V}$ is the volume, $J$ is the interlayer exchange coupling between the two easy-plane FMs, and $A$ and $K$ represent the spin stiffness and easy-plane anisotropy of the FM, respectively.
$\textbf{s}_{i} (\textbf{r}) = S_{i} (\textbf{r})/S$ denotes the spin orientation of each easy-plane FM, 
$i = \pm$ denotes the top (bottom) FM, and $S$ is the saturated spin density.
For simplicity, we assume the device to be symmetric around the center of the NM layer so that the layer
thicknesses, interfaces, and material parameters in the top half of the SSJO 
are identical to those in the bottom half. 
In Eq. (\ref{eq_Hamiltonian}), the easy-plane anisotropy ($K>0$) ensures that the energy of each FM is independent of the in-plane magnetization direction (indicating $U(1)$ symmetry). 
The energy is minimized by the in-plane spin configuration, $s_{i}^z=0$ with a fixed in-plane magnetization angle ($\phi_{i} =0$).

For the in-plane spin configuration, the last term in Eq. (\ref{eq_Hamiltonian}) becomes an oscillatory function of the relative angle of the in-plane magnetization. 
The interlayer exchange coupling $J$ can be AFM ($J>0$) or FM ($J<0$). 
This depends on the material and geometrical parameters~\cite{parkin_oscillations_1990}.
For $J>0$, spins in the top and bottom easy-plane FMs point in 
opposite directions cancelling the dipolar interactions.
It has been shown that dipolar interactions in easy plane FMs can break the easy-plane U(1) symmetry which destroys spin superfluidity in long distance spin transport~\cite{skarsvag_spin_2015}.
Hereafter, for simplicity, we assume an AFM type interlayer exchange coupling $J>0$ and neglect dipolar interactions.

The long wavelength magnetization dynamics of the device heterostructure can be captured by the Landau-Lifshitz-Gilbert (LLG) equation,
\begin{eqnarray}
\label{LLG}
\hbar \dot{\textbf{s}}_{i} = {} & -\textbf{s}_{i} \times (A\nabla^{2} \textbf{s}_{i}+K ({s}_{i}^z  \hat{\textbf{z}})  + J  \textbf{s}_{-i})\\ \nonumber 
& +\alpha \textbf{s}_{i} \times \dot{\textbf{s}}_{i}  +\bm{\tau}_{{s, i}}  ,
\end{eqnarray}
where $\dot{\textbf{s}}_{i}$ denotes the time derivative of $\textbf{s}_{i}$, $\alpha$ is the damping constant and
$\bm{\tau}_{{s, i}}$ describes the spin torque and spin pumping effect at the HM/FM interfaces.
To induce the spin oscillations, a spin current with polarization perpendicular to the easy-plane, can be used to produce a spin transfer torque on the in-plane magnetization. 
This spin current is generated by driving an electrical current in the x-direction in the HM contacts via the SHE.
Decomposing, $\textbf{s}_{i} = (\sqrt{1-({s}_{i}^z)^2} \cos\phi_{i}, \sqrt{1-({s}_{i}^z)^2} \sin\phi_{i}, {s}_{i}^z)$, the dynamics of both FMs can be expressed in terms of canonically conjugate amplitude and phase variables, ${s}_{i}^z$ and $\phi_{i}$.
For small variation of the out-of-plane magnetization, ${s}_{i}^z \ll 1$, 
the LLG equation can be expanded to the lowest order in ${s}_{i}^z$ and $\phi_{i}$,
\begin{eqnarray}
\nonumber
\hbar \dot{s}_{i}^z &=&  J \sin(\phi_{i} - \phi_{-i}) - \hbar \alpha \dot{\phi_{i}} + \tau_{i,s}^{z}, \\
\hbar \dot{\phi}_{i} &=& K {s}_{i}^z + J {s}_{-i}^z+ \hbar \alpha \dot{s}_{i}^z,
\label{Eq_magnetization_dynamics}
\end{eqnarray}
where we assume that $\tau_{s,i}$ is perpendicular to the easy plane. 
We take the single domain approximation of magnetization dynamics across the junction, and hereafter neglect the spatial dependence of the fields and set $\nabla^2 \textbf{s}_{i} = 0$.

Defining $\phi= \phi_{i} - \phi_{-i}$ as the relative phase, $\tau^{z}_{s} = \tau^{z}_{s,i} - \tau^{z}_{s,-i}$ as the total spin torque across the junction and $n = {s}_{i}^z - {s}_{-i}^z$ as the relative out-of-plane magnetization, the equation of motion can be reduced into two equations that describe the dynamics of the relative coupled variables,
\begin{eqnarray}
\label{eq_Josephson}
\hbar \dot{n} &=& 2 J \sin(\phi) -  \hbar \alpha \dot{\phi} + \tau^z_{s}, \\ \nonumber
\hbar \dot{\phi} &=& (K-J) n +\hbar \alpha \dot{n}.
\end{eqnarray}
For zero damping and zero spin torque ($\alpha = 0, \tau^z_{s}=0$), the above equations resemble the $\pi$-phase Josephson junction of weak link superconductors with characteristic frequency $\omega_{0} = \sqrt{2J (K - J)} / \hbar$~\cite{liu_spin-josephson_2016}.

\begin{figure}
\begin{center}
\includegraphics[width=3.5in,height=1.8in]{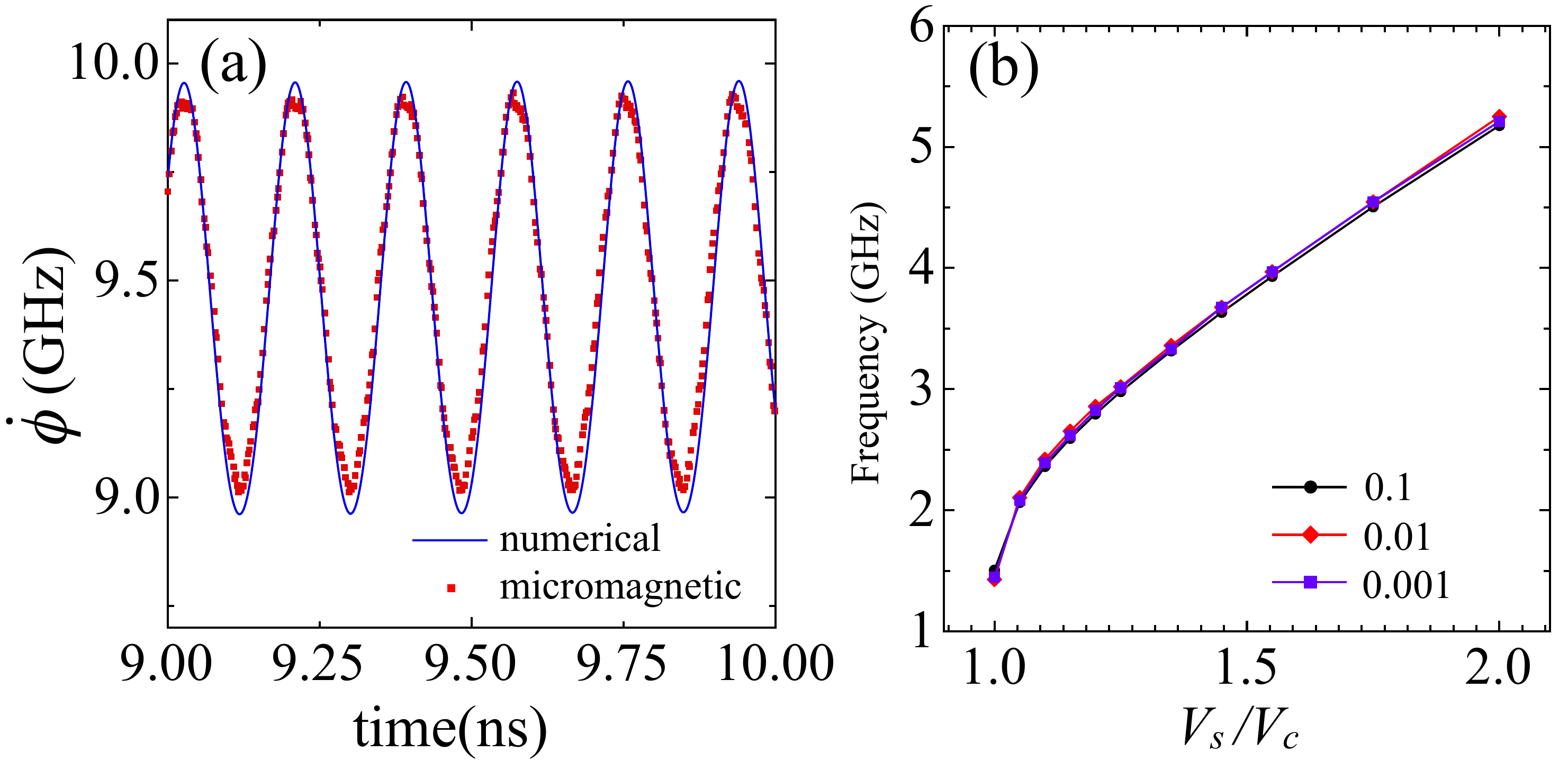}
\caption{(a) The steady state time dynamics of $\dot{\phi}$ from both numerical and micromagnetic simulations. (b)  Frequency as a function of the spin chemical potential for three different damping constants in the underdamped regime.}
\label{fig:frequency}
\end{center}
\end{figure}

\section{Dynamics with dc input}
When biased as shown in Fig.~\ref{fig:structure}, the two HMs are connected in a parallel circuit configuration with currents flowing in the same direction (x-direction).
The current flowing through the HM contacts generates a spin chemical potential, with spins perpendicular to the easy plane.
To determine the electric and magnetic dynamics in the circuit, we solve for their dynamics self-consistently in the device heterostructure.
The spin torque at the HM/FM interface can be expressed as,
\begin{equation}
\bm{\tau}_{s,i} = \frac{g_s}{4\pi} [  {\textbf{s}_{i} \times (\bm{\mu}_{0,i} \times \textbf{s} _{i})} - \hbar \textbf{s}_{i} \times\dot{\textbf{s}_{i}} ].
\label{Eq_torque}
\end{equation}
where $\bm{\mu_{0,i}} = \mu \hat{z}$,  denotes the non-equilibrium spin accumulation at the top and bottom interface.
$g_{s} = \frac{\mathcal{A} G^{R}_{\uparrow \downarrow}} {NS}$ is the effective spin mixing conductance, where $\mathcal{A}$ is the interface area, $G^{R}_{\uparrow \downarrow}$ is the real part of the spin mixing conductance and N is the total number of spins.
We assume the spin mixing conductance to be purely real through out this paper. 
The first term in Eq.~\ref{Eq_torque} is the spin torque exerted due to the injected spin current. 
The second term is the reciprocal spin pumping effect~\cite{tserkovnyak_enhanced_2002} due to the precession of the magnetization in the FMs.
This term must be included to satisfy Onsager's reciprocity relations.

%
With the circuit set up above, we have $V_{s,i} = -V_{s,-i} = |V_s|/2$, where the spin chemical potential $ V_{s} = g_{s}\mu/(4\pi)$.
Inserting this relationship into Eq. (\ref{Eq_magnetization_dynamics}) and eliminating $n$, we get the equation of motion for $\phi$~\cite{liu_spin-josephson_2016},
\begin{equation}
\label{Eq_RCSJ}
(1+ \tilde{\alpha} \alpha)\ddot{\phi}  + \frac{\hbar \tilde{\alpha} {\omega_0}^2}{2J}  \dot{\phi} -  {\omega_0}^2 \sin({\phi}) = \frac{{\omega_0}^2}{J}  V_s,
\end{equation} 
where $\tilde{\alpha} = \alpha + g_{s}/(4 \pi)$ is the enhanced damping. 
This equation is the same as the RCSJ model for superconducting Josephson junctions with an effective Stewart-McCumber parameter 
$\beta = 2J(1+\alpha \tilde{\alpha})/(\tilde{\alpha}^2(K - J))$~\cite{mccumber_effect_1968, stewart_currentvoltage_1968}.
Because of the Gilbert damping a critical spin chemical potential $V_{c}$, which is related to the current density in the HM contacts, is required to excite a magnetization oscillation.
In the strong damping regime ($\beta \ll 1$), the critical spin chemical potential required to induce a persistent  $\dot{\phi}$ oscillation is $V_{c} = 2J$.
In the intermediate damping regime ($\beta \sim 1$), the critical spin chemical potential can be estimated as $V_{c} = 2\tilde{\alpha} \sqrt{2JK}$, which depends on the damping, the interlayer exchange coupling and the FM easy-plane.
To determine the magnetization dynamics, we perform both micromagnetic simulations 
(red dots in Fig. {\ref{fig:frequency}} (b)) and numerical calculations of the RCSJ model 
(solid line in Fig. {\ref{fig:frequency}} (b)) for the SSJ junction.
The micromagnetic simulations were performed on two exchange coupled easy-plane FMs with dimensions 200 nm $\times$ 200 nm $\times$ 5 nm.
The saturation magnetization in micromagnetic simulations is $M_s=140$ kA/m and the parameters for
Eq. (\ref{eq_Hamiltonian}) are $A = 12.5$ meV, $K=0.25$ meV, $J=1$ $\mu$eV, and $\alpha = 0.01$.
In steady-state, $\dot{\phi}$ oscillates around a finite time-averaged value. 
These oscillations correspond to $2 \pi$ rotation of the relative in-plane magnetization.
Comparison of the micromagnetic simulations validates our Hamiltonian and the RCSJ model.

Above the critical voltage $V_{c}$, the oscillation frequency depends on the applied voltage $V_s$.
To investigate the relationship, we plot the oscillation frequency as a function of the normalized spin chemical potential $V =V_s/V_c$ for three different damping constants in Fig.~\ref{fig:frequency}.
Within this damping regime, the frequency has a quasi linear relationship with $V$ which is independent of damping.
In the strong damping regime ($\beta \ll1$), one can analytically solve Eq.~(\ref{Eq_RCSJ}) and obtain a time-dependent solution for $\dot{\phi}$ with an oscillation frequency $\omega= \sqrt{{V_s}^2 - 4J^2}/(h \tilde{\alpha})$.
This is consistent with our numerical calculations of the RCSJ model.

\begin{figure}
\begin{center}
\includegraphics[width=2.6in,height=3.6in]{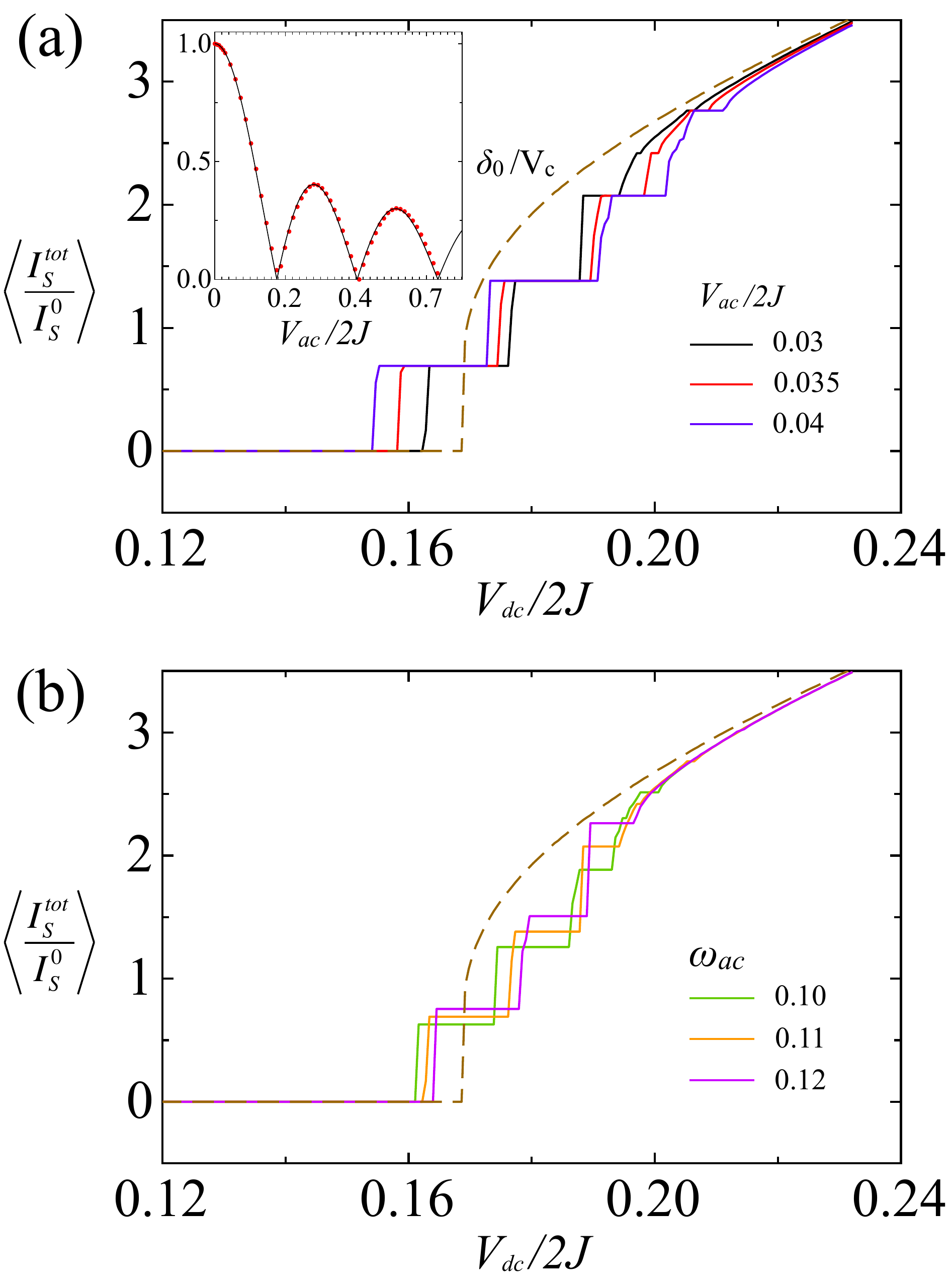}
\caption{Numerical results of the mode locking for: (a) different values of $V_{ac}$ and fixed frequency $\omega_{ac}=0.11$ and (b) different frequencies with fixed amplitude $V_{ac}/2J = 0.03$. The brown dashed line represents the case with d.c. input only. The damping constant is $\alpha =0.1$. Inset: The width of the zeroth step $\delta_0$ as a function of the ac input $V_{ac}$. Red dots denote the numerical results and the solid line is a fit to the $|J_0|$ Bessel function.}
\label{fig:steps}
\end{center}
\end{figure}

\section{Dynamics with ac input}
The spin oscillations also exhibit Shapiro step like dynamics in the presence of an ac input \cite{shapiro_josephson_1963}.
An ac spin chemical potential of the form of ${{V}_{s,i}} (t)=-{V}_{s,-i} (t) = V_{d.c.} + V_{ac} \sin(\omega_{ac} t)$ can be induced by 
an ac electric current in the HMs.
In a simple model, when $\omega_{ac}$ is an integer multiple of $\omega_0$, mode-locking of the input signal and the magnetization dynamics can occur.
This results in a Shapiro step like behavior in the time averaged pumped spin current $I_s$ within the HM.
The dynamics of the system with ac input are determined by solving Eq. (\ref{Eq_RCSJ}) with a time dependent source $V_{s}$.

In order to explore the role of $\omega_{ac}$ and $V_{ac}$, we calculate the time averaged $I_s$ as a function of $V_{s}$.
The results for different $V_{ac}$ and $\omega_{ac}$ are shown in Fig.~\ref{fig:steps}(a) and Fig.~\ref{fig:steps}(b), respectively.
The normalized spin current $I^{tot}_{s}/I^{0}_{s}$ is plotted where $I^{0}_{s} = \hbar g_{s} \omega_{0} / 4\pi$  is the characteristic spin current associated with the SSJ junction. 
When $V_{ac} = 0$, the time averaged $I_s$-$V_s$ shows a non-linear relationship with $V_{d.c.}$~\cite{liu_spin-josephson_2016}.
After increasing $V_{ac}$ to a finite value, the $I_s$-$V_s$ curve shows several steps, in the averaged spin current.

At each step, the oscillation frequency and maximum oscillation amplitude depends on $V_{dc}$.
The step position in the time averaged $I_{s}$ has a complicated dependence on $V_{dc}$, $\omega_{ac}$ and $V_{ac}$.
Both $V_{ac}$ and $\omega_{ac}$ strongly influence the d.c. critical value and the step width, whereas the step height only depends on $\omega_{ac}$.
As shown in Fig.~\ref{fig:steps}, finite $V_{ac}$ can reduce the critical value required for a persistent oscillation.
Furthermore, the width of the $n^{th}$ step $\delta_{n}$ as a function of $V_{ac}/2J$ is proportional 
to the Bessel function of the first kind $|J_n (V_{ac}/2J)|$, 
which is characteristic of the Shapiro steps \cite{shapiro_effect_1964} in superconducting Josephson junctions.
The magnetization dynamics of the SSJOs generate a novel magnetoresistance that we discuss in the next section.

\begin{figure}
\begin{center}
\includegraphics[width=2.8in,height=1.8in]{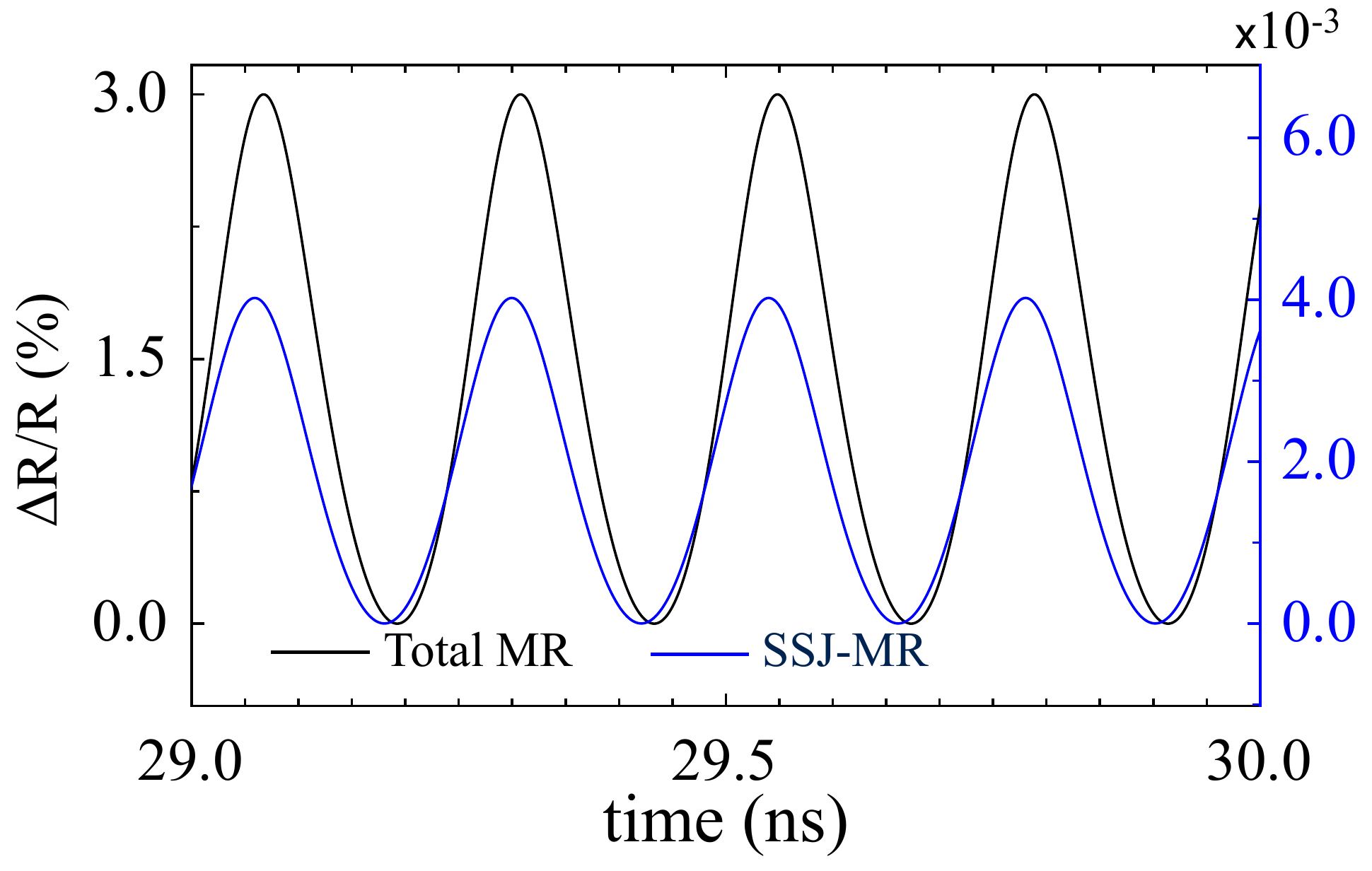}
\caption{Steady-state oscillation of the magnetoresistance of the SSJ-MR and the total MR. 
The values for the total MR are given by the left vertical axis,
and the values for the SSJ-MR are given by the right vertical axis.
Note the scale factor above the right vertial axis.
The damping constant is 0.01 and the spin chemical potential $V_s = 1.6V_c$.
The magnetoresistance is calculated using the parameters given in the main text. 
}
\label{fig:MR}
\end{center}
\end{figure}

\section{Output Power}
Spin pumping generates an additional electromotive force, 
\begin{equation}
{\Delta} \bm{E}_{i} = i \frac{{{\hbar g_{s} \theta_{\mbox {\tiny SH}}}}}{2e d_{\mbox{\tiny HM}}} (\textbf{s}_{i}  \times \dot{\textbf{s}}_{i}) \times \hat{\bf y} 
\label{Eq_Delta_E}
\end{equation}
where $i = \pm$ denotes the top and bottom HM contacts with thickness $d_{HM}$.
$\theta_{SH}$ is the effective spin Hall angle that takes into account the thickness effect of the HM. 
The current in the HM contacts is modified to ${\bm j} = ({\bm {E}} + \Delta {\bm {E}})/ \rho$, where the first term is the applied field, 
$\rho$ is the resistivity and $j$ is the electrical current density.
Eq. (\ref{Eq_Delta_E}) indicates that the magnetization dynamics generate an internal electromotive force which can be detected as magnetoresistance.
Therefore, the SSJ oscillations can be detected by purely electrical means.

One can apply Eq. (\ref{Eq_Delta_E}) to the device geometry in Fig.~\ref{fig:structure} and determine the MR response.
We restrict our analysis to the $x$-component of $ \Delta \bm{E}$, since the $y$-component of $ \Delta \bm{E}$ is small. 
For ${s}_{i}^z \ll 1$, we find $(\textbf{s}_{i}  \times \dot{\textbf{s}}_{i}) \times \hat{\bf y} = \dot{\phi_i} \hat{\bf x}$.
According to Eq. (\ref{Eq_Delta_E}), $\Delta \bm{E} = \bm{E}_{i} - \bm{E}_{-i}$ is directly proportional to $\dot{\phi}$.
Thus, the dynamics of $\dot{\phi}$ gives $\Delta \bm{E}$ a time-dependent contribution, which results in an effective MR of the circuit with an oscillation frequency $\omega$.
The SSJ-MR with a d.c. electric current source in the HM contacts is plotted in the Fig.{~\ref{fig:MR}}.
The magnitude of the SSJ-MR signal is primarily determined by the spin-Hall angle $\theta_{SH}$.
%
%

In metallic SSJOs, which consist of two easy-plane FM metals, additional contributions due to the in-plane giant magnetoresistance (GMR) dominate the MR signal~\cite{mcguire_anisotropic_1975,campbell_hall_1970, binasch_enhanced_1989, baibich_giant_1988, hutten_evolution_1999}.
%
%
The GMR contribution, $R_{GMR} = R_0 + \Delta R_{GMR} (1-cos(\phi))/2$, can be large
%
owing to the global $2\pi$ precession of the in-plane magnetization.
To estimate the total MR, we consider a device with an area of 200 nm by 50 nm in the $x-z$ plane as defined in Fig. \ref{fig:structure}. 
The junction consists of two 3 nm thick metallic easy-plane FMs separated by a 2 nm non-magnetic metal 
and sandwiched by two 5 nm Pt contacts, corresponding to the geometry shown in Fig. \ref{fig:structure}.
Using the following parameters, 
effective spin Hall angle of Pt $\theta_{\mbox{{\tiny SH}}} = 0.1$, sheet resistance of Pt $R^{\mbox{\tiny Pt}}_s=30 \Omega$, 
interfacial spin mixing conductance $G^{R}_{\uparrow \downarrow} = 5 \times 10^{18} \mbox{m}^{-2}$, GMR ratio of 15\%, 
and sheet resistance of the junction, $R = 40$ $\Omega$~\cite{egelhoff_optimizing_1996}, 
we calculate the resistance change for both the GMR and SSJ-MR.
The total MR for metallic SSJOs with a d.c. source, plotted in Fig{~\ref{fig:MR}}, is three orders of magnitude larger than the MR in insulating SSJOs. 
Here, due to the SSJ junction dimensions and large resistance of the FM layers, 
the contribution due to the anisotropic magnetoresistance (AMR) can be ignored.

MR signals are commonly employed to estimate the output power of a spin torque nano-oscillator.
For the SSJOs, the $2 \pi$ magnetization dynamics within the easy-plane could potentially enhance the output power of the proposed device.
%
%
In metallic SSJ junctions the full $2 \pi$ precession angle provides access to the maximum values of the GMR for a 
given $\Delta R$.
The output power for a $50 \; \Omega$ load ranges from 200 nW to 280 nW, depending on the GMR ratio, roughly an order of magnitude larger than other spin Hall oscillators.

Shapiro steps can still be detected for metallic SSJ junctions in the time averaged $\Delta R$.
Since the amplitude of $\Delta R_{\rm GMR}$ does not depend on $\dot{\phi}$, the time averaged contribution due to GMR is zero.
Thus, even in a metallic system, the non-linear $\Delta R$-$V_s$ and the Shapiro-step behavior still persists and is not buried by the large GMR effect.

Easy-plane anisotropy can be realized in FM thin films, but injecting spin polarization normal to the easy-plane via 
the SHE is challenging.
One scheme is to create an easy-plane perpendicular to the sample plane ($x$-$z$ plane) as shown in Fig.~\ref{fig:structure}. 
It has been shown that such an easy-plane can be engineered in Co/Ni by 
combining the shape anisotropy and an easy-axis magnetic anisotropy~\cite{easy_plane_oscillator}.
In a sample geometry with a specific aspect ratio, such as a nanowire, 
an easy-axis lies along the nanowire direction ($x$-axis) due to the shape anisotropy.
By carefully tuning an easy-axis anisotropy perpendicular to the nanowire ($y$-axis), 
the out-of-plane anistropy induced by the dipole interaction can be fully canceled 
leaving an easy-plane perpendicular to the $z$-axis.

\section{Conclusion}
A new type of spin nano-oscillator based on a SSJ effect is proposed and analyzed.
A spin chemical potential difference across the junction drives planar magnetization rotation.
This spin oscillation is mediated by a spin superfluid mode and directly related to the phase difference between the two FMs.
The oscillation frequency is tuned by the interlayer exchange and the spin chemical potential.
The output power is enhanced by the GMR effect in metallic SSJ junctions.
The $2\pi$ precession angle of the spin superfluid mode maximizes the GMR effect, 
thus opening an alternate route towards building high power spin oscillators.
The multi-state mode-locking behavior exhibited by the spin Shapiro steps may also be exploited for applications in neuromorphic computing.
As an experimental characterization method, 
electrical measurements of SSJ junctions can provide additional signatures of spin superfluidity.
\\

\noindent
{\em Acknowledgements}:
This work was supported as part of the Spins and Heat in Nanoscale Electronic Systems (SHINES) an 
Energy Frontier Research Center funded by the U.S. Department of Energy, Office of Science, 
Basic Energy Sciences under Award No. DE-SC0012670.

%

\end{document}